\documentstyle[12pt]{article}
\newcommand{\beq}{\begin{equation}}
\newcommand{\eeq}{\end{equation}}
\newcommand{\api}{\frac{\alpha_s}{\pi}}

\newcommand{\apiq}{\frac{\alpha_s(Q^2)}{\pi}}
\newcommand{\apimu}{\frac{\alpha_s(\mu^2)}{\pi}}
\newcommand{\apimuo}{\frac{\alpha_s(\mu_0^2)}{\pi}}

\newcommand{\apimt}{\frac{\alpha_s(M_{\tau})}{\pi}}
\newcommand{\ba}{\begin{array}} 
\newcommand{\ea}{\end{array}} 
\newcommand{\ds}{\displaystyle} 
\newcommand{\as}{\alpha_s}
\newcommand{\gm}{\gamma_m}
\newcommand{\G}{\Gamma}
\newcommand{\g}{\gamma}

\newcommand{\gvam}{\gamma^{VV/AA}_m}

\newcommand{\dmu}{\mu^2\frac{d}{d\mu^2}}
\newcommand{\msbar}{\overline{\mbox{MS}}}
\newcommand{\kmu}{\dsp\frac{(m_i(\mu^2)\mp m_j(\mu^2))^2}{Q^2}}
\newcommand{\kq}{\dsp\frac{(m_i(Q^2)\mp m_j(Q^2))^2}{Q^2}}

\newcommand{\kqp}{\dsp\frac{(m_i(Q^2)\pm m_j(Q^2))^2}{Q^2}}
\newcommand{\pva}{\Pi^{(0)}_{V/A}}
\newcommand{\pvaij}{\Pi^{(0)}_{V/A;ij}}

\newcommand{\pvaijtr}{\Pi^{(0+1)}_{V/A;ij}}

\newcommand{\dvaijv}{\delta^{(D=4)}_{V/A;ij}}
\newcommand{\mtau}{M_{\tau}}
\newcommand{\kk}{\dsp\frac{1}{16\pi^2}}
\newcommand{\dsp}{\displaystyle}

\newcommand{\EQN}{\label}

\begin{document}

\begin{titlepage}
\noindent
%
%
%
%
\vspace{0.5cm}
\begin{center}
  \begin{Large}
  \begin{bf}
Mass Corrections to the  Tau Decay Rate\footnote{Corrected version
of the paper published in Z.~Phys. C59 (1993) 525}
  \\
  \end{bf}
  \end{Large}
%
%
  \vspace{0.8cm}
  \begin{large}
K.G.Chetyrkin\footnote{On leave from Institute for Nuclear Research
of the Russian Academy of Sciences, Moscow, 117312, Russia.}\\[3mm]
    Institut f{\"u}r Theoretische Teilchenphysik\\
    Universit{\"a}t Karlsruhe\\[2mm]
    D-76128 Karlsruhe 1, Germany\\[5mm]
 A.Kwiatkowski    \\[3mm]
    Physik-Department\\
    Technische Universit{\"a}t M{\"u}nchen\\[2mm]
    D-85747 Garching, Germany\\
  \end{large}
%
%
  \vspace{5.8cm}
  {\bf Abstract}
\end{center}
\begin{quotation}
\noindent
In this note radiative corrections to the total hadronic decay rate of the
$\tau$-lepton are studied employing perturbative QCD and the operator product
expansion.
We calculate
 quadratic quark mass corrections
to the decay rate ration $R_{\tau}$ to the order
 ${\cal O}(\alpha_s^2 m^2)$ and find that they contribute
appreciably to the Cabbibo supressed decay modes of the $\tau$-lepton.
We also discuss corrections of mass dimension $D=4$, where
we emphasize the need of a suitable choice of the renormalization scale
of the quark and gluon condensates.
\end{quotation}
\end{titlepage}

\section{Introduction}

\renewcommand{\arraystretch}{2}                             %
Within the three lepton generations which  to our present knowledge constitute
the leptonic sector of the Standard Model, the $\tau$-lepton is the only
particle decaying into  a semihadronic final state.
As  was pointed out 
 \cite{SchTra84,Bra88,NarPic88,Bra89,BraNarPic92,LeDPic92a,LuoMar92,LeDPic92b}
some time ago,  the methods of perturbative QCD
can be applied to estimate the decay rate ratio
\beq    \EQN{1}
R_{\tau} = \frac{\G(\tau\rightarrow\nu_{\tau} {\rm hadrons})}
          {\G(\tau\rightarrow\nu_{\tau}e \bar{\nu}_e)}
.\eeq
Work has also concentrated on  nonperturbative effects
 \cite{SchTra84,Bra88,NarPic88,BraNarPic92,LeDPic92b}
as well as on electroweak corrections \cite{MarSir88,BraLi90} to this quantity.

In order to calculate QCD corrections to $R_{\tau}$  we consider
the 2-point correlators  for the vector
 ($j^V_{\mu,ij}=\bar{q_i}\gamma_{\mu}q_j$)
and the axial vector
($j^A_{\mu,ij}=\bar{q_i}\gamma_{\mu}\g_5 q_j$) currents ($i,j=u,d,s$):
\beq    \EQN{2}
 \Pi^{V/A}_{\mu\nu,ij}(q)= i \int dx e^{iqx}
\langle
T[j^{V/A}_{\mu,ij}(x)j^{V/A\dagger}_{\nu,ij}(0)] \rangle
\eeq
As usual the correlation functions may be decomposed into a transversal and a
longitudinal part
\beq   \EQN{3}
 \Pi^{V/A}_{\mu\nu}(q)= (-g_{\mu\nu}q^2+q_{\mu}q_{\nu}) \Pi^{(1)}_{V/A}(q^2)
                       + q_{\mu}q_{\nu} \Pi^{(0)}_{V/A}(q^2),
\eeq
where the spectral functions $\Pi^{(0)}_{V/A}(q^2),\Pi^{(1)}_{V/A}(q^2)$
correspond to hadronic final states
with respective angular momenta $J=0, J=1$ in the hadronic rest frame.
The hadronic decay rate of the $\tau$-lepton is obtained by integrating
the absorptive parts of the spectral functions with respect to the
invariant hadronic mass:
\beq      \EQN{4}
R_{\tau}=12\pi\int_0^{M_{\tau}^2}\frac{ds}{M_{\tau}^2}
\left(1-\frac{s}{M_{\tau}^2}\right)^2
\left[\left(1+2\frac{s}{M_{\tau}^2}\right){\rm Im}\Pi^{(1)}(s+i\epsilon)
+{\rm Im}\Pi^{(0)}(s+i\epsilon)\right],
\eeq
where
\beq    \EQN{5}
\Pi^{(J)}=|V_{ud}|^2(\Pi^{(J)}_{ud,V}+\Pi^{(J)}_{ud,A})
         + |V_{us}|^2(\Pi^{(J)}_{us,V}+\Pi^{(J)}_{us,A}).
\eeq
Due to the analyticity of the spectral functions in the complex $s$-plane,
which is cut along the real positive $s$-axis, $R_{\tau}$ can be expressed
as the contour integral along a circle C of radius
$|s|=M_{\tau}^2$:
\beq \EQN{6}
R_{\tau}=6i\pi\int_{|s|=M_{\tau}^2}\frac{ds}{M_{\tau}^2}
\left(1-\frac{s}{M_{\tau}^2}\right)^2
\left[\left(1+2\frac{s}{M_{\tau}^2}\right)\Pi^{(0+1)}(s)
-2\frac{s}{M_{\tau}^2}\Pi^{(0)}(s)\right]
\eeq
 We have used in this equation the combination
 $\Pi^{(0+1)}(q^2)=\Pi^{(0)}(q^2)+\Pi^{(1)}(q^2)$.

$R_{\tau}$ may be expressed as the sum of different contributions corresponding
to Cabbibo suppressed or allowed decay modes, vector or axial vector
contributions and the mass dimension of the corrections:
\beq \EQN{7}
R_{\tau} = R_{\tau,V} + R_{\tau,A} + R_{\tau,S}
\eeq
with
\beq \EQN{8}
\ba{ll} \ds
R_{V} =
& \ds
\frac{3}{2}|V_{ud}|^2 \left( 1 + \delta^{(0)} + \sum_{D=2,4,\dots}
\delta^{(D)}_{V,ud} \right),  \\
R_{A} =
& \ds
\frac{3}{2}|V_{ud}|^2 \left( 1 + \delta^{(0)} + \sum_{D=2,4,\dots}
\delta^{(D)}_{A,ud} \right),  \\
R_{S} =
& \ds
3|V_{us}|^2 \left( 1 + \delta^{(0)} + \sum_{D=2,4,\dots}
\delta^{(D)}_{us} \right).
\ea      
\eeq
Here $D$ indicates the mass dimension of the fractional corrections
$\delta^{(D)}_{ij,V/A}$ and $\delta^{(D)}_{ij}$  denotes the 
average of the vector and the axial vector contributions:
 $\delta^{(D)}_{ij,}=(\delta^{(D)}_{ij,V}+\delta^{(D)}_{ij,A})/2$.

In the literature results for perturbative
QCD corrections  to current-current correlators have been published
by several groups. 
For  the electromagnetic current correlator corrections were calculated
in the massless limit  up to ${\cal O}(\as^2)$
\cite{Che79} and  ${\cal O}(\as^3)$ \cite{GorKatLar91}, whereas
quadratic   mass corrections to this quantity 
were given to second ${\cal O}(\as^2)$ \cite{GorKatLar86} and third 
${\cal O}(\as^3)$ order \cite{CheKue90}.
For the axial current correlator mass corrections are known from two loop
\cite{Bro,BroGen84,Gen89} and three loop calculations
 \cite{CheKueKwi92} for flavour non singlet type diagrams.
 Flavour singlet contributions 
to the axial correlator involving purely gluonic intermediate states 
were studied  in the heavy top limit in \cite{KniKue90a,KniKue90b}
and for the massive case in \cite{CheKwi92}. 
Finally second order massless corrections  for
the scalar current correlator can be found  in  \cite{GorKatLarSur90}.

A detailed analysis of QCD corrections for the $\tau$ decay into hadrons 
has been performed in \cite{BraNarPic92}, where much attention
focussed on  nonperturbative
contributions to corrections of higher dimension.

The aim of this letter is twofold.
First we extend the analysis
 of ref.\ \cite{BraNarPic92} and calculate the order ${\cal O}(\as^2)$ term
of  the (dimension $D=2$)
longitudinal spectral function $ \Pi^{(0)}_{V/A}(q^2)$ in its
power expansion with respect to $\alpha_s$. 
This analysis is done  in section 2.

\noindent
The second part of this note is contained in
 section 3 where we consider corrections
of mass dimension $D=4$ to $R_{\tau}$.
We estimate their size where we use a scale $\hat{\mu}$ 
for the quark condensates
which corresponds to the energy scale of the process under consideration. 
Our choice $\hat{\mu}^2=M_{\tau}^2$ seems to be more appropriate to us 
than  $\hat{\mu}^2=\infty$ as used in \cite{BraNarPic92}. 

\noindent
Our numerical results are discussed in section 4.
We finally list  some
formulae in the appendix.

\section{Dimension $D=2$ Corrections}
The way to derive the quadratic mass correction to the
longitudinal spectral function $\pvaij$
is based on the
knowledge of the anomalous dimensions $\g_m^{VV/AA}$ of the vector
and the axial vector correlators as defined in  \cite{CheKueKwi92}
\beq \EQN{9}
\dmu \Pi^{V/A}_{\mu\nu}=\frac{q_{\mu}q_{\nu}-g_{\mu\nu}q^2}{16\pi^2}
\g_q^{VV/AA}(\as)
+(m_i\mp m_j)^2 \frac{g_{\mu\nu}}{16\pi^2}
\g_m^{VV/AA}(\as).
\eeq
Up to order ${\cal O}(\as^2)$ the anomalous dimension $\g_m^{VV/AA}$
is given by (with $f$ denoting the number of quark flavours)
\beq      \EQN{10}
 \gvam = 6\left( 1 + \frac{5}{3}\apimu
+ \left(\apimu\right)^2\left[\frac{455}{72}-\frac{1}{3}f-\frac{1}{2}\zeta(3)
\right]\right)
\eeq
and governs the renormalization group equation for the longitudinal spectral
function
\beq     \EQN{11}
\dmu \pvaij = -\kmu\gvam \frac{1}{16\pi^2}.
\eeq
In order to get  $\pva$   the RGE eq.(\ref{9})must be integrated.
 The solution reads
\beq     \EQN{12}    \ba{ll} \dsp
\pvaij = & \dsp \kk  \kmu  \exp\left[2\int^{\apimu}_{\apimuo}dA^{\prime}
\frac{\gm}{\beta}\right]   \\ & \dsp
\hphantom{\kk}\cdot
\int^{\apimu}_{\apimuo}dA^{\prime}\frac{-\gvam}{\beta}
\exp\left[-2\int^{A^{\prime}}_{\apimuo}dA^{\prime\prime}
\frac{\gm}{\beta}\right] + C(\mu_0^2) ,
\ea \eeq
where the $\beta$-function  and the quark mass anomalous dimension $\gm$
are given in the appendix.
 The integration constant $C(\mu_0^2)$ may be fixed by a specific choice
for the arbitrary scale $\mu_0^2$ which we choose to be $\mu_0^2=Q^2$. This 
leads to 
\beq  \EQN{13}
C(Q^2) = \frac{3}{8\pi^2}\kq\left\{-2
+\apiq\left[4\zeta(3)-\frac{131}{12}\right]\right\}     .
\eeq
With the  relation between the $\mu^2$-dependent mass of the
$\msbar$-scheme and the $Q^2$-dependent running mass (see appendix)
\beq \EQN{14}
m^2(\mu^2) = m^2(\mu_0^2)
\exp\left[-2\int^{\apimu}_{\apimuo}dA^{\prime}
\frac{\gm}{\beta}\right]
\eeq
we  cast $\pvaij$ into the following form
\beq \ba{rl}  \EQN{15}
\pvaij =&  -\kk \kq \left\{
\exp\left[2\int^{\apiq}dA^{\prime}
\frac{\gm}{\beta}\right] \right.   \\[4ex] &\dsp
\hphantom{\kk}\left. \cdot
\int^{\apiq}dA^{\prime}\frac{-\gvam}{\beta}
\exp\left[-2\int^{A^{\prime}}dA^{\prime\prime}
\frac{\gm}{\beta}\right]-C(Q^2)   \right\}      \\[4ex]
&
  +\kk \kmu
\exp\left[2\int^{\apimu}dA^{\prime}
\frac{\gm}{\beta}\right]    \\[4ex] & \dsp
\hphantom{\kk}\cdot
\int^{\apimu}dA^{\prime}\frac{\gvam}{\beta}
\exp\left[-2\int^{A^{\prime}}dA^{\prime\prime}
\frac{\gm}{\beta}\right]
.\ea \eeq
Inserting the series expansion for $\beta$ and $\gm$ as given in the
appendix we obtain after integration
\beq  \ba{rl}    \EQN{16}
\pvaij = & \ds
\frac{3}{2\pi^2} \kq \left\{\left(\apiq\right)^{-1} - \frac{5}{2}\right.  \\
& \ds \hphantom{\frac{3}{2\pi^2} \kq }   \left.
+ \left[-\frac{21373}{4896}+\frac{75}{34}\zeta(3)\right] \apiq \right\}    \\
&\ds -\frac{3}{2\pi^2} \kmu \left\{\left(\apimu\right)^{-1} - 2  \right. \\
& \ds \hphantom{-\frac{3}{2\pi^2} \kmu }   \left.
+ \left[-\frac{8011}{4896}+\frac{41}{34}\zeta(3)\right] \apimu \right\}
\ea \eeq
The leading and next to leading terms of this expansion agree with
\cite{BraNarPic92} whereas the  coefficient of $\as$ is new.
The $\mu^2$-dependent
part  of eq.(\ref{16})
does not contribute to the hadronic $\tau$-decay rate.

\vspace{1ex}
Mass corrections
 of the transversal spectral function $\pvaijtr$  were calculated
to order ${\cal O}(\as^2m^2)$
for the electromagnetic correlation function in 
 \cite{GorKatLar86}. 
Recently one of us (K.~Ch.)  extended this calculation to the case of 
unequal masses. The obtained result reads
\beq
  \ba{rl}    \EQN{17}
\pvaijtr = &  \ds
-\frac{3}{8\pi^2} \kqp \left\{1 + \frac{8}{3}\apiq \right.  \\
&\ds \hphantom{-\frac{3}{8\pi^2} \kqp }   \left.
+ \left[\frac{17981}{432}+\frac{62}{27}\zeta(3)-\frac{1045}{54}\zeta(5)\right]
\left(\apiq\right)^2 \right\}    \\
&\ds -\frac{3}{8\pi^2} \kq  \left\{1 + 2 \apiq \right. \\
&\ds \hphantom{-\frac{3}{8\pi^2} \kq }   \left.
+ \left[\frac{4351}{144}+\frac{13}{3}\zeta(3)-\frac{115}{6}\zeta(5)\right]
\left(\apiq\right)^2 \right\}    \\
& \dsp +\frac{1}{12\pi^2}\left(\apiq\right)^2
(32-24\zeta(3))\sum_k\frac{m_k(Q^2)}{Q^2}    
.\ea \eeq
Even for the decay of the $\tau$-lepton into non strange quarks
is a dependence on $m_s$ introduced by the   second order contributions
due to the  s-quark circulating in  a fermion loop.
The term proportional to $\sum_k m_k(Q^2)/Q^2$ was first computed in 
\cite{BerWet81} 
For the  vector
correlator with $m_i=m_j$ the result eq.(\ref{17})
exactly reproduces the one of 
\cite{GorKatLar86}. This means that a slightly different 
result (namely for the terms
proportional to $\zeta(3)$)  for the diagonal vector correlator given in
\cite{Sur89} seems to be wrong. 
For a detailed discussion we refer the reader to \cite{phys_report}.

\section{Dimension $D=4$ Corrections}
Besides the perturbative radiative corrections to $R_{\tau}$ also
nonperturbative QCD effects influence the hadronic $\tau$ decay rate.
The short distance operator product expansion (OPE) for the spectral functions
\beq \EQN{t1}
\Pi^{(J)}(-q^2) = \sum_{D=0,2,4,\cdots}\frac{1}{(-s)^{D/2}}
               \sum_{{\rm dim}{\cal O} = D} C^{(J)}(Q^2,\mu)
               \langle{\cal O}(\mu)\rangle
\eeq
may be used to take into account both perturbative and non perturbative
contributions. 
 For the fractional corrections $\delta^{(D)}_{V/A;ij}$
(see eq.(\ref{8}))
 we obtain
\beq \EQN{t2}
\ba{ll}
\delta^{(D)}_{V/A;ij}  =
& \dsp
\sum_{{\rm dim}{\cal O} = D}
\frac{\langle{\cal O}(\mu)\rangle}{\mtau^D}4\pi i\int_{|s|=\mtau^2}
\frac{ds}{\mtau^2}\left(\frac{-s}{\mtau^2}\right)^{-D/2}
\left(1-\frac{s}{\mtau^2}\right)^2
\\ & \ds
\hphantom{\sum_{{\rm dim}{\cal O} = D}\frac{<{\cal O}(\mu)>}{\mtau^D}}
\left[\left(1+2\frac{s}{\mtau^2}\right)C^{(0+1)}_{ij,V/A}(s,\mu)
-2\frac{s}{\mtau^2}C^{(0)}_{ij,V/A}(s,\mu)\right].
\ea 
\eeq
The local operators 
 in this expansion for the $D=4$ perturbative
corrections are the unit operator multiplied by a quartic product of 
quark masses and the vacuum expectation values of composite operators 
constructed with gluon and quark fields: 
$\langle GG(\mu)\rangle,\langle m_i(\mu)\bar{\Psi}_j\Psi_j(\mu)\rangle$.
 The latter contain nonperturbative contributions 
\cite{ShiVaiZak79,ReiRubYaz85}
as well as 
mass logarithms of the form \cite{BroGen84,CheSpi88}
 $m^4 \as^n(\mu)\ln^k(m/\mu)$
and depend nontrivially on the renormalization scale $\mu$ via the 
corresponding renormalization group equations \cite{CheSpi88}.
As usual, setting the renormalization scale $\mu=Q$ in eq.(\ref{t1})
allows to absorb all logarithms $\ln \mu^2/Q^2$ appearing in the 
coefficient functions $C^{(J)}$ into the running coupling constant.
This procedure also leads to an implicit $Q^2$-dependence of 
VEV's, which is not convenient for a numerical analysis. The common remedy
is to to solve the corresponding RG equations and express
 $\langle {\cal O}\rangle$ in terms of $\as(Q)$ and some RG invariant
combination not depending on $Q^2$.

As has been shown in \cite{CheSpi88} it is possible to construct
 linear
combinations of the operators  which are invariant with respect to an arbitrary
scale
 $\hat{\mu}$:
\beq  \ba{ll}    \EQN{t3}
\langle{\cal I}_G\rangle&\equiv  \ds
\left(1+\frac{16}{9}\frac{\as(\hat{\mu}^2)}{\pi}\right)
  \frac{\as(\hat{\mu}^2)}{\pi} \langle GG(\hat{\mu})\rangle \\
&\ds
-\frac{16}{9}\frac{\as(\hat{\mu}^2)}{\pi}
 \left(1+\frac{91}{24}\frac{\as(\hat{\mu}^2)}{\pi}\right)
\sum_k \langle m_k(\hat{\mu})\bar{\Psi}_k\Psi_k(\hat{\mu})\rangle \\
& \ds
-\frac{1}{3\pi^2} \left(1+\frac{4}{3}\apimu\right) \sum_k m_k^4(\hat{\mu}) , \\
\langle{\cal I}_{ij}\rangle
&\equiv \ds
m_i(\hat{\mu})\langle\bar{\Psi}_j\Psi_j(\hat{\mu})\rangle \\
&\ds
+\frac{3}{7\pi\as(\hat{\mu})} \left(1-\frac{53}{24}\frac{\as(\hat{\mu}^2)}{\pi}
\right)m_i(\hat{\mu})m_j^3(\hat{\mu})
,\ea \eeq
These combinations were intensively used in the analysis of \cite{BraNarPic92}
under the names of the gluon condensate $\langle \frac{\as}{\pi}GG\rangle$
and the quark condensate $\langle m_i\bar{\Psi}_j \Psi_j\rangle$ 
respectively. Of course the choice is by no means unique. Moreover,
it directly leads to an $1/\as$ enhancement factor in the dimension $D=4$
correction to the ratio $R_{\tau}$ (see eq.(3.10) in \cite{BraNarPic92}).
This results in a partial cancellation between quartic mass corrections and
those coming from the quark condensates, which in the final analysis causes
some partial loss of accuracy of theoretical predictions \cite{BraNarPic92}).
 
In addition one can easily check that 
\beq \EQN{t3a}
\ba{ll} \dsp
\langle {\cal I}_G\rangle = &  \dsp
\lim_{\hat{\mu}\to\infty} \frac{\as(\hat{\mu})}{\pi}\langle
 GG(\hat{\mu})\rangle, 
\\ \dsp
\langle {\cal I}_{ij}\rangle = &  \dsp
\lim_{\hat{\mu}\to\infty} \langle m_i(\hat{\mu})\bar{\Psi}_j
\Psi_j(\hat{\mu})\rangle
.\ea\eeq
Thus the choice of eq.(\ref{t3}) as RG invariant vacuum condensates features a
quite high normalization scale  for operators involved in a problem 
with a typical momentum transfer of about 1 GeV. No wonder that this causes an
artificially large perturbative mass correction of order ${\cal O}(m^4)$ 
to various 2-point correlators as was found in \cite{Bro}.
 
Keeping all this in mind we have chosen just the very quark and gluon
condensates normalized at the ``natural'' scale $\hat{\mu}=M_{\tau}$
as our RG invariant condensates:
\beq \EQN{t3b}
\ba{ll} \dsp
\langle \frac{\as}{\pi}GG\rangle = &  \dsp
 \langle\frac{\as(M_{\tau})}{\pi} GG(\hat{\mu}=M_{\tau})\rangle, 
\\ \dsp
\langle m_i\bar{\Psi}_j\Psi_j\rangle = &  \dsp
\langle m_i(M_{\tau})\bar{\Psi}_j\Psi_j(\hat{\mu}=M_{\tau})\rangle
.\ea\eeq
Using the RG invariance property of the combinations eq.(\ref{t3}) 
it is a straightforward exercise to express the $Q^2$-dependent
VEV's in terms of our RG invariants and the running $\as(Q),
m(Q)$.

The result for the fractional corrections  after performing the contour
integral reads  
\beq   \EQN{t4}
\ba{rl}
\ds
\dvaijv \cdot M_{\tau}^4 =
& \ds
\frac{11}{4}\pi^2\left(\apimt\right)^2 \langle \api GG\rangle  \\
& \ds
+16\pi^2 \left[1+\frac{9}{2}\left(\apimt\right)^2 \right]
\langle (m_i\mp m_j)(\bar{\Psi}_i\Psi_i\mp \bar{\Psi}_j\Psi_j) \rangle \\
& \ds
-18\pi^2\left(\apimt\right)^2
 \langle m_i\bar{\Psi}_i\Psi_i+m_j\bar{\Psi}_j\Psi_j \rangle            \\
& \ds
-8\pi^2\left(\apimt\right)^2   \sum_k\langle  m_k\bar{\Psi}_k\Psi_k \rangle
     \\
& \ds
- \frac{84}{7}   [m_i(M_{\tau})\mp m_j(M_{\tau})]
                [m_i^3(M_{\tau})\mp m_j^3(M_{\tau})] \\
& \ds
\pm 6  m_i(M_{\tau})m_j(M_{\tau})[m_i(M_{\tau})\mp m_j(M_{\tau})]^2   \\
& \ds
+  36  m_i^2(M_{\tau})m_j^2(M_{\tau})
\ea \eeq
For some terms the leading coefficient  is of order  ${\cal O}(\as^2)$
due to the fact
that a term of given order of the transversal
spectral function is contributing only in  higher orders to the
fractional correction and order ${\cal O}(\as^0)$ terms may eventually
integrate to zero.
We have neglected terms of order $\as m^4$ in eq.(\ref{t4}).
No  enhanced term  proportional to the 
inverse power of the coupling constant occurs as was
the case 
for the choice of reference \cite{BraNarPic92}.

\section{Numerical Discussion}
The numerical discussion in this section is based on the parameters
(quark masses, condensates etc.) as they are given in the appendix.
Depending on the value of $\Lambda_{QCD}$ the running coupling constant
$\as(M_{\tau}^2)$ ranges between $0.16$ and $0.44$ and is thus sufficiently
small for the perturbation expansion to be meaningful.
The quadratic mass corrections $\delta^{(D=2)}_{V/A;ij}$ obtained from the
contour integration of the spectral functions eqs.(\ref{16},\ref{17})
 are collected in Table 1\footnote{We thank K. Maltman
\cite{Mal98} for pointing
out an error in the original version of this paper.}.
 One observes that corrections from nonstrange
decays are negligible, whereas the mass of the strange quark affects the ratio
$R_S$ for strange decays by $-20$\% for an intermediate $\as(M_{\tau}^2)=0.3$.
When the total hadronic ratio $R_{\tau}$ is considered, this rather large
strange mass contribution is reduced by the Cabbibo supression factor
$|V_{us}|^2$ to $-0.8$\% ($V_{ud}=0.9753,\;V_{us}=0.221$).
The influence of the second order correction 
 depends of course on the
value  of $\Lambda_{QCD}$.  Even for nonstrange decays strange
quark  mass effects
are present at order ${\cal O}(\as^2)$ due to a virtual strange quark loop.
They are quite comparable in size to the leading mass corrections of order
$(m_u\pm m_d)^2$ and 
enter with opposite sign.
Due to their increasing size for larger $\as$ the 
corrections $\delta^{(D=2)}_{V/A;ud}$ show little dependence on 
$\Lambda_{QCD}$.
For strange decays of the $\tau$-lepton 
we recall the numerical values of the 
coefficients entering the spectral functions (in the limit of vanishing
masses of the light quarks)
\beq  \EQN{n1}
\ba{ll} \ds
\Pi^{(0)}_{V/A;us} = &  \ds
\frac{3}{2\pi^2} \frac{m_s^2(Q^2)}{Q^2}
 \left\{\left(\apiq\right)^{-1} - \frac{5}{2}
- 1.714\apiq \right\}    \\
&\ds -\frac{3}{2\pi^2}  \frac{m_s^2(\mu^2)}{Q^2}
 \left\{\left(\apimu\right)^{-1} - 2 - 1.867\apimu \right\}
,\ea
\eeq

\beq
    \EQN{n2}
\Pi^{(0+1)}_{V/A;us} = 
-\frac{3}{4\pi^2} \frac{m_s^2(Q^2)}{Q^2} 
\left\{ 1+\frac{7}{3}\apiq +19.583\left(\apiq\right)^2\right\}
\eeq
and the fractional correction
\beq    \EQN{n3}
\delta^{(2)}_{V/A;us} = -8 \frac{m_s^2(M_{\tau})}{M_{\tau}^2}
  \left\{1 + \frac{16}{3}\apimt + 46.002 \left(\apimt\right)^2 \right\}   
.\eeq          
Second order contributions to  $\delta^{(D=2)}$ originate not only from
order ${\cal O}(\as^2)$ terms of the spectral functions, but are
also induced by lower order terms. 

In Table 2   the corrections of mass dimension $D=4$ are shown.
The change of scale from $\hat{\mu}=\infty$ to  $\hat{\mu}=M_{\tau}$
significantly  affects only the Cabbibo supressed vector contribution.
The small corrections $\delta^{(D=2)}_{V;us}$ in the analysis of
\cite{BraNarPic92} were due to a numerical cancellation between the leading
quark condensate and an $1/\as$-enhanced mass term $m_s^4$. This resulted in
a large relative uncertainty for this correction. When the condensates are 
defined at the scale $\hat{\mu}=M_{\tau}$ a similar compensation of 
numerically large terms does not occur due to the absence of the 
$1/\as$-enhanced mass term.
 Compared to the corresponding numbers 
of \cite{BraNarPic92} the fractional corrections   $\delta^{(D=2)}_{V;us}$
are increased by an order of magnitude which reduces the relative uncertainty
considerably. In addition they do not change their sign at large
values of $\Lambda_{QCD}$. The strange $D=4$ corrections contribute only little
(namely -3\%) to the strange decay ratio $R_S$ and are negligible 
when the total $\tau$-decay ratio $R_{\tau}$ is considered.  

We now include all corrections to 
 present  the values for the  separate contributions 
as well as for the total  hadronic $\tau$-decay rate in Table 3. 
Here electroweak as well as corrections of higher mass dimension 
have been taken into account from  
 \cite{BraNarPic92}, where it has been
pointed out that the biggest nonperturbative corrections 
to $R_{\tau}$ in fact come
from dimension $D=6$ condensates. Our estimation of the 
corresponding uncertainty on
$R_{\tau}$ is more conservative than  in \cite{BraNarPic92}, because
the large cancellation between the vector and the axial vector 
contribution is based on the validity of vacuum saturation approximation
and may not necessarily be translated into a similar compensation for the 
uncertainties of the separate contributions. 
Numerically  $R_\tau$ is dominated by 
purely perturbative contributions of zero mass  dimensions which 
survive in the massless limit. According to
\cite{Piv92}
a  proper summation of the effects of analytical continuation from 
space-like momenta to time-like ones is a necessity  for these terms.  
At the moment it is not quite clear for us  how essential  these 
effects are  for power suppressed contributions that we are dealing with.  We 
hope to return to this and related problems in future publications.

To conclude, we have studied power suppressed
perturbative and nonperturbative QCD
corrections to the semileptonic decay rate of the $\tau$-lepton.
Strange quark mass effects contribute considerably 
to the decay modes with strangeness content and still reach the percent
level for the total hadronic decay rate.  
 Second order ${\cal O}(\as^2)$
corrections introduce  a strange mass dependence even for non strange
decays.
We also have studied corrections of mass dimension $D=4$, where we discussed
an appropriate choice for the renormalization scale of the quark and gluon
condensates. 
In view of the accuracy of QCD predictions for the 
  decay rate  of the $\tau$-lepton into hadrons, semileptonic
$\tau$ decays remain an important and interesting tool for testing
QCD.

\vspace{5ex}
{\bf Table 1:} $\delta^{(D=2)}_{ij,V/A}$ fractional corrections

\nopagebreak
\vspace{2ex}
\begin{tabular}{|r|r||r|r||r|r||}
\hline
$\Lambda_{QCD}$/MeV & $\as(M_{\tau})$ &$\delta^{(D=2)}_{ud,V}\cdot 10^3$
& $\delta^{(D=2)}_{ud,A}\cdot 10^3$ & $\delta^{(D=2)}_{us,V}$
   & $\delta^{(D=2)}_{us,A}$  \\
\hline
52& 0.16&-0.20 $\pm$ 0.04& -0.37 $\pm$ 0.07
& -0.074 $\pm$ 0.017& -0.077$ \pm$  0.017     \\
\hline
308& 0.30&-0.36 $\pm$ 0.12& -0.87 $\pm$   0.20
& -0.197 $\pm$ 0.044 & -0.206 $\pm$ 0.045     \\
\hline
547& 0.44&-0.41 $\pm$ 0.29& -1.60 $\pm$ 0.46
& -0.418 $\pm$ 0.094& -0.439 $\pm$    0.096 \\ \hline
\end{tabular}

\vspace{5ex}
{\bf Table 2:} $\delta^{(D=4)}_{ij,V/A}$ fractional corrections

\nopagebreak
\vspace{2ex}
\begin{tabular}{|r|r||r|r||r|r||}
\hline
$\Lambda_{QCD}$/MeV & $\as(M_{\tau})$ &$\delta^{(D=4)}_{ud,V}\cdot 10^3$
&$ \delta^{(D=4)}_{ud,A}\cdot 10^3$ & $\delta^{(D=4)}_{us,V}$
   & $\delta^{(D=4)}_{us,A}$  \\
\hline
52& 0.16& 0.17 $\pm$ 0.13& -5.00 $\pm$ 0.61
&  0.011 $\pm$ 0.005& -0.048 $\pm$  0.007     \\
\hline
308& 0.30& 0.60 $\pm$ 0.27& -4.68 $\pm$   0.67
&  0.010 $\pm$ 0.005 & -0.050 $\pm$ 0.008    \\
\hline
547& 0.44& 1.30 $\pm$ 0.55& - 4.20 $\pm$ 0.83
&0.009  $\pm$ 0.005&-0.055 $\pm$    0.009 \\ \hline
\end{tabular}

\vspace{5ex}
{\bf Table 3:} Contributions to the hadronic $\tau$ decay rate $R_{\tau}$

\nopagebreak
\vspace{2ex}
\begin{tabular}{|r|r||r|r|r||r||}
\hline
$\Lambda_{QCD}$/MeV & $\as(M_{\tau})$ &$R_{V}$
&$ R_{A}$ & $R_{S}$
   & $R_{\tau}$  \\
\hline
52& 0.16& 1.59 $\pm$ 0.02& 1.49 $\pm$ 0.03
&  0.145 $\pm$ 0.003& 3.23 $\pm$  0.02     \\
\hline
308& 0.30& 1.73 $\pm$ 0.02& 1.63 $\pm$   0.03
&  0.140 $\pm$ 0.007 & 3.51 $\pm$ 0.05   \\
\hline
547& 0.44& 1.95 $\pm$ 0.08&  1.85 $\pm$ 0.08
&0.128  $\pm$ 0.016& 3.93 $\pm$    0.16 \\ \hline
\end{tabular}

\vspace{5ex}
{\bf Acknowledgments}

\noindent
We would like to thank A.Pich and J.H.K{\"u}hn for helpful discussions.
One of us (K.G.Ch.) is also grateful to M.Jamin, K.Schilcher and  
D.Pirjol for numerous fruitful discussions of mass corrections and 
their relation to vacuum condensates.
The work was supported by Bundesministerium f{\"u}r Forschung und
Technologie
(K.G.Ch., Project No. 055KA94P1).

\appendix
\section{Appendix}
The $\beta$-function and the quark mass anomalous dimension $\gm$ are defined
in the usual way
\beq  \EQN{a1}
\dmu \left( \frac{\as(\mu)}{\pi} \right) = \beta(\as) \equiv
-\sum_{i\geq0}\beta_i\left(\api\right)^{i+2}, \eeq
\beq    \EQN{a2}
\dmu \bar{m}(\mu) = - \bar{m}(\mu)\gm(\as) \equiv
-\bar{m}\sum_{i\geq0}\g_i\left(\api\right)^{i+1}. \eeq
Their expansion coefficients up to order ${\cal O}(\as^2)$ are well known
\cite{TarVlaZha80,Tar82} and read
\beq
\ba{ll}\dsp
\beta_0&=\dsp\left(11-\frac{2}{3}f\right)/4,  \  \
\beta_1=\left(102-\frac{38}{3}f\right)/16, \\[3ex] \dsp
\beta_2&=\dsp\left(\frac{2857}{2}-\frac{5033}{18}f+ 
\frac{325}{54}f^2\right)/64, 
\ea  \eeq \vspace{3ex}
\beq 
\ba{ll}\dsp 
\g_0&=\dsp 1, \  \ \ \g_1=\left(\frac{202}{3}-\frac{20}{9}f\right)/16,
\\[3ex] \dsp \g_2&=\dsp \left(1249 - 
\left[\frac{2216}{27}+\frac{160}{3}\zeta(3)\right] 
f-\frac{140}{81}f^2\right)/64.
\ea 
\eeq
The Riemann zeta function has the values 
$\zeta(3)=1.2020569, \zeta(5)=1.036927$.
The solution of eq.\ (\ref{a1}) is given by
 ($L\equiv \ln \mu^2/\Lambda^2_{\overline{MS}}$)
\beq
\frac{\as(\mu^2)}{\pi} = \frac{1}{\beta_0 L} \left[
1 - \frac{1}{\beta_0L}\frac{\beta_1\ln L}{\beta_0}
 + \frac{1}{\beta_0^2L^2} \left(\frac{\beta_1^2}{\beta_0^2}
(\ln^2L - 
\ln L - 1) + \frac{\beta_2}{\beta_0} \right) \right]
\eeq
while eq.\ (\ref{a2}) is solved by
\beq \ba{rl}\dsp
m(\mu^2) 
=& \dsp
\hat{m}\left(2\beta_0\frac{\as(\mu^2)}{\pi}\right)^{\g_0/\beta_0}
\left\{ 1+\apimu
        \left[\frac{\g_1}{\beta_0}-\frac{\beta_1\g_0}{\beta_0^2}\right]\right.
\\[3ex] &\dsp
  +\frac{1}{2} \left(\apimu\right)^2\left[
  \left[\frac{\g_1}{\beta_0}-\frac{\beta_1\g_0}{\beta_0^2}\right]^2
 \left.
+  \frac{\g_2}{\beta_0}-\frac{\beta_1\g_1}{\beta_0^2}
       -\frac{\beta_2\g_0}{\beta_0^2}+\frac{\beta_1^2\g_0}{\beta_0^3}
\right] \right\}
\ea\eeq    
For the renormalization invariant quark mass parameters we have taken the
values  \cite{Nar89,Dom}
\beq
\hat{m}_u = (8.7\pm 1.5)MeV, \;\;\;
\hat{m}_d = (15.4\pm 1.5)MeV, \;\;\;
\hat{m}_s = (270.\pm 30)MeV.
\eeq
As input value for the gluon condensate we have used \cite{Nar89}
\beq
\langle\as GG \rangle = (0.02 \pm 0.01)GeV^4,
\eeq
whereas the quark mass condensates are parametrized by \cite{Nar89}
$\langle m_i\bar{\Psi}_j\Psi_j\rangle  = -\hat{m}_i\hat{\mu}_j^3$ with
\beq
\hat{\mu}_u = \hat{\mu}_d = (189\pm 7) MeV, \;\;\;
\hat{\mu}_s = (160 \pm 10) MeV.
\eeq


\begin{thebibliography}{99}

\bibitem{SchTra84} 
K.Schilcher, M.D.Tran, Phys.\ Rev.\ D 29 (1984) 570.

\bibitem{Bra88}
E.Braaten, Phys.\ Rev.\ Lett.\ 53 (1988) 1606.

\bibitem{NarPic88}
S.Narison, A.Pich,  Phys.\  Lett.\ B 211 (1988) 183.

\bibitem{Bra89} 
E.Braaten, Phys.\ Rev.\ D 39 (1989) 1458.

\bibitem{BraNarPic92} 
E.Braaten, S.Narison, A.Pich, Nucl.\ Phys.\ B 373 (1992) 581.

\bibitem{LeDPic92a} 
F. Le\  Diberder, A.Pich, Phys.\ Lett.\ B 286 (1992) 147.

\bibitem{LuoMar92} 
M.Luo, W.J.Marciano, Preprint BNL-47187, 1992.               

\bibitem{LeDPic92b} 
F. Le\  Diberder, A.Pich, Phys.\ Lett.\ B 289 (1992) 165.

\bibitem{MarSir88} 
W.Marciano, A.Sirlin, Phys.\ Rev.\ Lett.\ 61 (1988) 1815.

\bibitem{BraLi90} 
E.Braaten, C.S.Li, Phys.\ Rev.\ D 42  (1990) 3888.

\bibitem{Che79}
K.G.Chetyrkin, A.L.Kataev, F.V.Tkachov,  Phys.\ Lett.\ B 85  (1979) 277;\\
M.Dine, J.Sapirstein, Phys.\ Rev.\ Lett. 43 (1979) 668; \\
W.Celmaster, R.J.Gonsalves, Phys.\ Rev.\ Lett. 44 (1980) 560.

\bibitem{GorKatLar91}
S.G.Gorishny, A.L.Kataev, S.A.Larin, Phys.\ Lett. 259 (1991) 144;\\
L.R.Surguladze, M.A.Samuel,  Phys. Rev. Lett. 66 (1991) 560, 2416 (erratum) \\
L.R.Surguladze, M.A.Samuel,  Phys. Rev. D 44 (1991) 1602.
                                                        
\bibitem{GorKatLar86}
S.G.Gorishny, A.L.Kataev, S.A.Larin, Nuovo Cimento 92 (1986) 119. 

\bibitem{CheKue90}
K.G.Chetyrkin, J.H.K{\"u}hn, Phys.\ Lett.\ B 248 (1990) 359.       

\bibitem{Bro}
C.Becchi, S.Narisson, E. de Rafael, 
and F.J.Yndurain, Z. Phys. C -- Particles and Fields  8 (1981) 335,     \\
D.J.Broadhurst, Phys.\  Lett. 101B (1981) 423.

\bibitem{BroGen84}
D.J.Broadhurst and S.C. Generalis,
Open University preprint OUT-4102-12 (1984) (unpublished).

\bibitem{Gen89}
S.C.Generalis, J. Phys. G15 (1989) L225, \\  
S.C.Generalis, J. Phys. G16 (1990) L117.          

\bibitem{CheKueKwi92}
K.G.Chetyrkin, J.H.K{\"u}hn, A.Kwiatkowski, Phys.\ Lett.\ B 282 (1992) 221.       

\bibitem{KniKue90a}
B.Kniehl, J.H.K{\"u}hn, Phys.\ Lett.\ B 224 (1990) 229.
     
\bibitem{KniKue90b}
B.Kniehl, J.H.K{\"u}hn, Nucl.\ Phys.\ B 329 (1990) 547.
     
\bibitem{CheKwi92}
K.G.Chetyrkin, A.Kwiatkowski, Preprint TTP92-39, 1992.     

\bibitem{GorKatLarSur90} 
S.G.Gorishny, A.L.Kataev, S.A.Larin, L.R.Surguladze,
 Mod.\ Phys.\ Lett. A  5 (1990) 2703.



\bibitem{phys_report}K.G.~Chetyrkin, J.H.~K{\"u}hn and A.~Kwiatkowski, 
Phys.~Rep.\   277 (1996) 189.   

   

\bibitem{BerWet81}
W.Bernreuther, W.Wetzel, Z. Phys. C -- Particles and Fields 11 (1981) 113.

\bibitem{Sur89}
L.R.Surguladze, INR Preprint P-0639, Moscow, 1989.  

\bibitem{ShiVaiZak79} 
M.A.Shifman, A.L.Vainshtein, V.I.Zakharov, Nucl.\ Phys.\ B 147 (1979)
385, 448, 519.

\bibitem{ReiRubYaz85} 
L.J.Reinders, H.Rubinstein, S.Yazaki, Phys.\ Rep.\ 127 (1985) .

\bibitem{CheSpi88} 
K.Chetyrkin, K.G.Spiridonov, Sov.\ J.\ Nucl.\ Phys. 47 (1988) 3.
 
\bibitem{Mal98}      
K.~Maltman, Preprint hep-ph/9804298. 
 
\bibitem{Piv92}      
A.A.Pivovarov,  Z.Phys. C -- Particles and Fields 53 (1992) 461.

\bibitem{TarVlaZha80} 
O.V.Tarasov, A.A.Vladimirov, A.Yu.Zharkov, Phys.\ Lett.\ B 93 (1980)429.

\bibitem{Tar82} 
O.V.Tarasov, preprint JINR P2-82-900 (1982).

\bibitem{Nar89} 
S.Narison, ``QCD Spectral Sum Rules'' (Lecture Notes in Physics, Vol. 26),
World Scientific, Singapore 1989.

\bibitem{Dom}      
C.A.Dominguez, E.\ de Rafael, Ann.\ Phys.\ 174 (1987) 372; \\
C.A.Dominguez, C. van Gend, N.Paver, Phys.\ Lett.\ B 253 (1990) 1.

\end{thebibliography}
\end{document}